\documentclass[useAMS,usenatbib]{mn2e}
\usepackage{times}
\usepackage{graphicx}
\usepackage{subfigure}
\usepackage{psfrag}
\usepackage{amsmath}
\usepackage{amssymb}
\def\reff@jnl#1{{\rm#1\/}}
\def\aj{\reff@jnl{AJ}}                 % Astronomical Journal
\def\araa{\reff@jnl{ARA\&A}}           % Annual Review of Astron and Astrophys
\def\apj{\reff@jnl{ApJ}}               % Astrophysical Journal
\def\apjl{\reff@jnl{ApJ}}              % Astrophysical Journal, Letters
\def\apjs{\reff@jnl{ApJS}}             % Astrophysical Journal, Supplement
\def\ao{\reff@jnl{Appl.Optics}}        % Applied Optics
\def\apss{\reff@jnl{Ap\&SS}}           % Astrophysics and Space Science
\def\aap{\reff@jnl{A\&A}}              % Astronomy and Astrophysics
\def\aapr{\reff@jnl{A\&A~Rev.}}        % Astronomy and Astrophysics Reviews
\def\aaps{\reff@jnl{A\&AS}}            % Astronomy and Astrophysics, Supplement
\def\azh{\reff@jnl{AZh}}               % Astronomicheskii Zhurnal
\def\baas{\reff@jnl{BAAS}}             % Bulletin of the AAS
\def\jrasc{\reff@jnl{JRASC}}           % Journal of the RAS of Canada
\def\memras{\reff@jnl{MmRAS}}          % Memoirs of the RAS
\def\mnras{\reff@jnl{MNRAS}}           % Monthly Notices of the RAS
\def\pra{\reff@jnl{Phys.Rev.A}}        % Physical Review A: General Physics
\def\prb{\reff@jnl{Phys.Rev.B}}        % Physical Review B: Solid State
\def\prc{\reff@jnl{Phys.Rev.C}}        % Physical Review C
\def\prd{\reff@jnl{Phys.Rev.D}}        % Physical Review D
\def\prl{\reff@jnl{Phys.Rev.Lett}}     % Physical Review Letters
\def\pasp{\reff@jnl{PASP}}             % Publications of the ASP
\def\pasj{\reff@jnl{PASJ}}             % Publications of the ASJ
\def\qjras{\reff@jnl{QJRAS}}           % Quarterly Journal of the RAS
\def\skytel{\reff@jnl{S\&T}}           % Sky and Telescope
\def\solphys{\reff@jnl{Solar~Phys.}}   % Solar Physics
\def\sovast{\reff@jnl{Soviet~Ast.}}    % Soviet Astronomy
\def\ssr{\reff@jnl{Space~Sci.Rev.}}    % Space Science Reviews
\def\zap{\reff@jnl{ZAp}}               % Zeitschrift fuer Astrophysik
\def\nat{\reff@jnl{Nature}}            % Nature
\title[]{Comment on ``Bayesian evidence: can we beat MultiNest using traditional MCMC methods", by Rutger van Haasteren (arXiv:0911.2150)}  
\author[F. Feroz, M.P. Hobson and R. Trotta]
{F. Feroz$^1$,\thanks{E-mail: f.feroz@mrao.cam.ac.uk} M.P. Hobson$^1$ and  R. Trotta$^2$\\
1 Astrophysics Group, Cavendish Laboratory, JJ Thomson Avenue, Cambridge CB3 0HE, UK\\
2 Astrophysics Group, Imperial College, Blackett Laboratory, Prince Consort Road, London SW7 2AZ, UK\\
}

\date{Accepted ---. Received ---; in original form \today}
\pagerange{\pageref{firstpage}--\pageref{lastpage}}
\pubyear{2008}

\begin{document}
\label{firstpage}
\maketitle

\begin{abstract}
In arXiv:0911.2150, Rutger van Haasteren seeks to criticize the nested sampling algorithm for Bayesian data analysis in general and its {\sc MultiNest} implementation in particular. He introduces a new method for evidence evaluation based on the idea of Voronoi tessellation and requiring samples from the posterior distribution obtained through MCMC based methods. He compares its accuracy and efficiency with {\sc MultiNest}, concluding that it outperforms {\sc MultiNest} in several cases. This comparison is completely unfair since the proposed method can not perform the complete Bayesian data analysis including posterior exploration and evidence evaluation on its own while {\sc MultiNest} allows one to perform Bayesian data analysis end to end. Furthermore, their criticism of nested sampling (and in turn {\sc MultiNest}) is based on a few conceptual misunderstandings of the algorithm. Here we seek to set the record straight.
\end{abstract}


%%%%%%%%%%%%%%%%%%%%%%%%%%%%%%%%%%%%%%%%%%%%%%%%%%%%%%%%%
\section{Introduction}\label{sec:intro}
%%%%%%%%%%%%%%%%%%%%%%%%%%%%%%%%%%%%%%%%%%%%%%%%%%%%%%%%%
In a recent paper (\citealt{vanHaasteren:2009yg}), Rutger van Haasteren has criticized the {\sc MultiNest} algorithm for Bayesian analysis and suggested another method to calculate the Bayesian evidence with the claim that it significantly outperforms {\sc MultiNest} both in terms of accuracy and computational accuracy. Our aim in this short note is to highlight the conceptual misunderstandings and the false premise on which the comparison is based.

The outline of the paper is as follows. In Sec. \ref{sec:bayesian} we give an introduction to Bayesian inference and describe the nested sampling algorithm and its implementation in {\sc MultiNest} package in Sec.~\ref{sec:multinest}. In Sec.~\ref{sec:critique} we give an account of the conceptual misunderstandings in \citealt{vanHaasteren:2009yg} which are the basis of the attack mounted by the author on {\sc MultiNest}. Finally, our conclusions are presented in Sec.~\ref{sec:conclusions}.

%%%%%%%%%%%%%%%%%%%%%%%%%%%%%%%%%%%%%%%%%%%%%%%%%%%%%%%%%
\section{Bayesian Inference}\label{sec:bayesian}
%%%%%%%%%%%%%%%%%%%%%%%%%%%%%%%%%%%%%%%%%%%%%%%%%%%%%%%%%
Bayesian inference methods provide a consistent approach both to the estimation of a set of parameters~$\mathbf{\Theta}$ in a model (or hypothesis) $H$ for the data $\mathbf{D}$ and to the evaluation of the relative merits of different models for the data (see~\cite{Trotta:2008qt} for a review of Bayesian methods in cosmology and astrophysics). Bayes' theorem states that
\begin{equation} \Pr(\mathbf{\Theta}|\mathbf{D}, H) =
\frac{\Pr(\mathbf{D}|\,\mathbf{\Theta},H)\Pr(\mathbf{\Theta}|H)}{\Pr(\mathbf{D}|H)},
\label{eq:bayes}
\end{equation}
where $\Pr(\mathbf{\Theta}|\mathbf{D}, H) \equiv P(\mathbf{\Theta})$ is the posterior probability distribution of the parameters, $\Pr(\mathbf{D}|\mathbf{\Theta}, H) \equiv \mathcal {L}(\mathbf{\Theta})$ is the likelihood, $\Pr(\mathbf{\Theta}|H) \equiv \pi(\mathbf{\Theta})$ is the prior distribution, and $\Pr(\mathbf{D}|H) \equiv \mathcal{Z}$ is the Bayesian evidence.

Bayesian evidence is the factor required to normalise the posterior over~$\mathbf{\Theta}$:
\begin{equation}
\mathcal{Z} = \int{\mathcal{L}(\mathbf{\Theta})\pi(\mathbf{\Theta})}d^D\mathbf{\Theta},
\label{eq:Z}
\end{equation}
where $D$ is the dimensionality of the parameter space. Since the Bayesian evidence is independent of the parameter values~$\mathbf{\Theta}$, it is usually ignored in parameter estimation problems and the posterior inferences are obtained by exploring the un--normalized posterior using standard MCMC sampling methods.

Bayesian parameter estimation has been used quite extensively in a variety of astronomical applications, including gravitational wave astronomy, although standard MCMC methods, such as the basic Metropolis--Hastings algorithm or the Hamiltonian sampling technique (see e.g. \citealt{MacKay}), can experience problems in sampling efficiently from a multi--modal posterior distribution or one with large (curving) degeneracies between parameters. Moreover, MCMC methods often require careful tuning of the proposal distribution to sample efficiently, and testing for convergence can be problematic.

In order to select between two models $H_{0}$ and $H_{1}$ one needs to compare their respective posterior probabilities given the observed data set $\mathbf{D}$, as follows:
\begin{equation}
\frac{\Pr(H_{1}|\mathbf{D})}{\Pr(H_{0}|\mathbf{D})}
=\frac{\Pr(\mathbf{D}|H_{1})\Pr(H_{1})}{\Pr(\mathbf{D}|H_{0})\Pr(H_{0})}
=\frac{\mathcal{Z}_1}{\mathcal{Z}_0}\frac{\Pr(H_{1})}{\Pr(H_{0})},
\label{eq:model_select}
\end{equation}
where $\Pr(H_{1})/\Pr(H_{0})$ is the prior probability ratio for the two models, which can often be set to unity but occasionally requires further consideration (see, for example, \citealt{2009MNRAS.398.2049F,2008arXiv0810.0781F} for the cases where the prior probability ratio should not be set to unity) and the ratio of the evidences $\frac{\mathcal{Z}_1}{\mathcal{Z}_0}$ is called the Bayes factor between the two models. It can be seen from Eq.~(\ref{eq:model_select}) that the Bayesian evidence plays a central role in Bayesian model selection. As the average of likelihood over the prior, the evidence automatically implements Occam's razor: a simpler theory which agrees well enough with the empirical evidence is preferred. A more complicated theory will only have a higher evidence if it is significantly better at explaining the data than a simpler theory~(e.g., \cite{Liddle:2004nh,Trotta:2005ar}).

The evaluation of the Bayesian evidence involves the multidimensional integral (Eq.~\ref{eq:Z}) and thus presents a challenging numerical task. Standard techniques like thermodynamic integration \cite{Ruanaidh} are usually extremely computationally expensive, which makes evidence evaluation typically at least an order of magnitude more costly than parameter estimation. Recently, a estimation of the evidence using population Monte Carlo techniques has been successfully implemented in the cosmological context~\cite{Kilbinger:2009by}. Some fast approximate methods have been used for evidence evaluation, such as treating the posterior as a multivariate Gaussian centred at its peak (see, for example, \citealt{Hobson02}), but this approximation is clearly a poor one for highly non-Gaussian and multi--modal posteriors. A computationally cheap and accurate method is the Savage-Dickey density ratio~\cite{Trotta:2005ar,Trotta:2006ww,Vardanyan:2009ft}, which however can only be used to compute the Bayes factor between nested models. Various alternative information criteria for model selection are discussed in \cite{Liddle:2004nh,Liddle07}, but the evidence remains the preferred method.

%%%%%%%%%%%%%%%%%%%%%%%%%%%%%%%%%%%%%%%%%%%%%%%%%%%%%%%%%
\section{Nested Sampling and the {\sc MultiNest} Algorithm}\label{sec:multinest}
%%%%%%%%%%%%%%%%%%%%%%%%%%%%%%%%%%%%%%%%%%%%%%%%%%%%%%%%%

%
\begin{figure}
\begin{center}
\subfigure[]{\includegraphics[width=0.3\columnwidth]{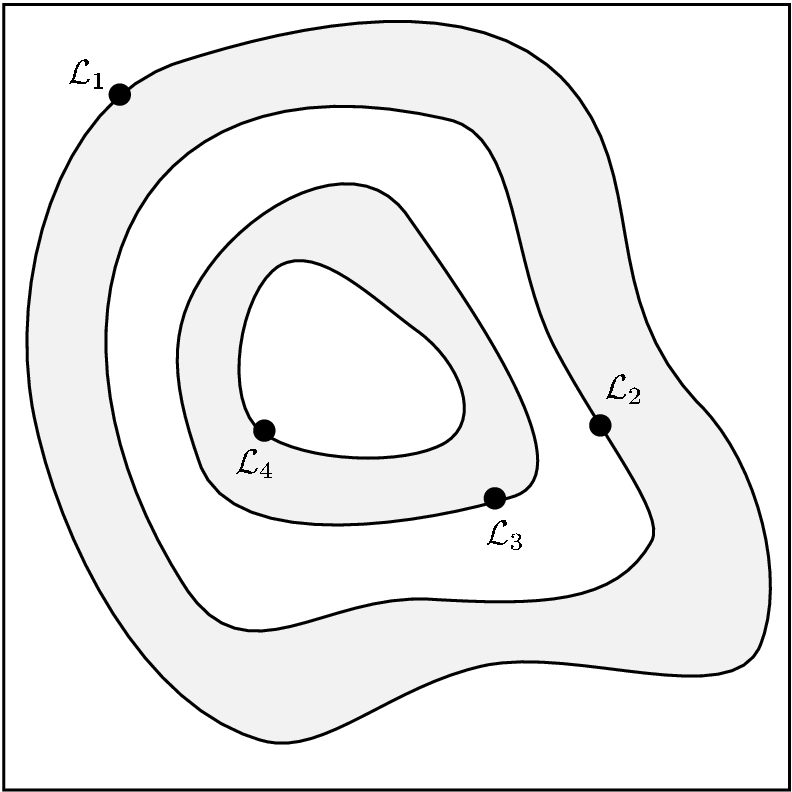}}\hspace{0.5cm}
\subfigure[]{\includegraphics[width=0.3\columnwidth]{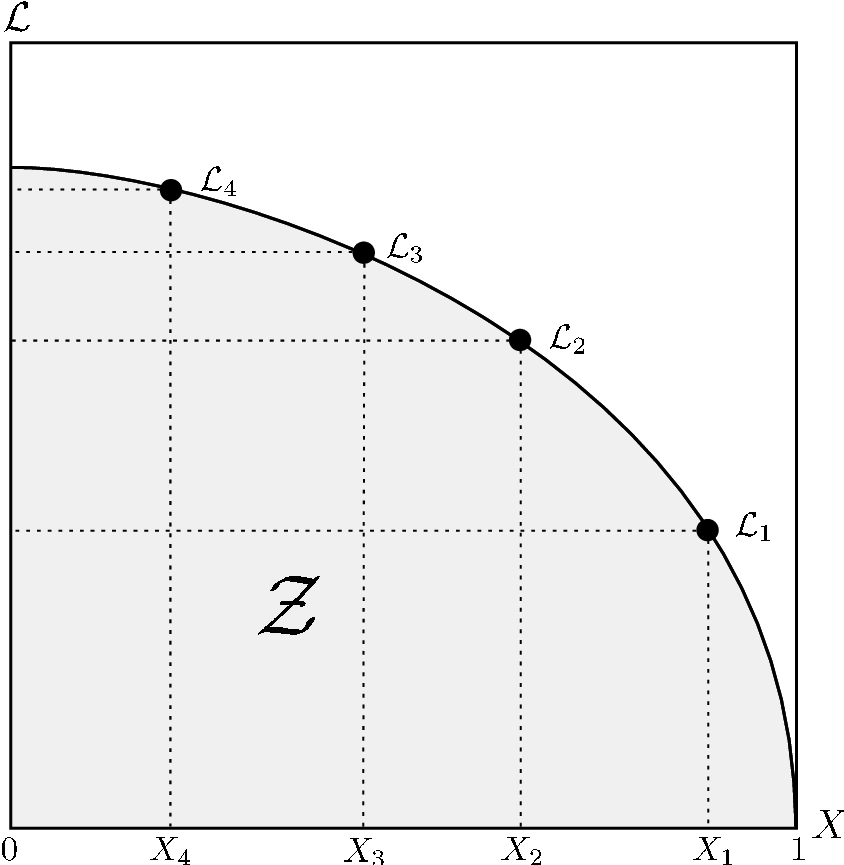}}
\caption{Cartoon illustrating (a) the posterior of a two dimensional problem; and (b) the transformed $\mathcal{L}(X)$  function where the prior volumes $X_{\rm i}$ are associated with each likelihood
$\mathcal{L}_{\rm i}$.}
\label{fig:NS}
\end{center}
\end{figure}

In this section we briefly review the Nested sampling algorithm and the {\sc MultiNest} implementation. Further details can be found in \cite{Skilling04, feroz08, multinest}. 

%%%%%%%%%%%%%%%%%%%%%%%%%%%%%%%%%%%%%%%%%%%%%%%%%%%%%%%%%
\subsection{Nested Sampling}\label{sec:nested}
%%%%%%%%%%%%%%%%%%%%%%%%%%%%%%%%%%%%%%%%%%%%%%%%%%%%%%%%%

Nested sampling~\cite{Skilling04} is a Monte Carlo method targetted at the efficient calculation of the evidence, but also produces posterior inferences as a by-product. It calculates the evidence by transforming the multi--dimensional evidence integral into a one--dimensional integral that is easy to evaluate numerically. This is accomplished by defining the prior volume $X$ as $dX = \pi(\mathbf{\Theta})d^D \mathbf{\Theta}$, so that
\begin{equation}
X(\lambda) = \int_{\mathcal{L}\left(\mathbf{\Theta}\right) > \lambda} \pi(\mathbf{\Theta}) d^D\mathbf{\Theta},
\label{eq:Xdef}
\end{equation}
where the integral extends over the region(s) of parameter space contained within the iso-likelihood contour $\mathcal{L}(\mathbf{\Theta}) = \lambda$. The evidence integral, Eq.~(\ref {eq:Z}), can then be written as
\begin{equation}
\mathcal{Z}=\int_{0}^{1}{\mathcal{L}(X)}dX,
\label{eq:nested}
\end{equation}
where $\mathcal{L}(X)$, the inverse of Eq.~(\ref{eq:Xdef}), is a  monotonically decreasing function of $X$.  Thus, if one can evaluate the likelihoods $\mathcal{L}_{\rm i}=\mathcal{L}(X {i})$, where $X_{\rm i}$ is a sequence of decreasing values,
\begin{equation}
0<X_{\rm M}<\cdots <X_{2}<X_{1}< X_{0}=1,
\end{equation}
as shown schematically in Fig.~\ref{fig:NS}, the evidence can be approximated numerically using standard quadrature methods as a weighted sum
\begin{equation}
\mathcal{Z}={\textstyle {\displaystyle \sum_{\rm i=1}^{M}}\mathcal{L}_{\rm i}w_{\rm i}},
\label{eq:NS_sum}
\end{equation}
where the weights $w_{\rm i}$ for the simple trapezium rule are given by $w_{\rm i}=\frac{1}{2}(X_{\rm i-1}-X_{\rm i+1})$. An example of a posterior in two dimensions and its associated function $\mathcal{L}(X)$ is shown in Fig.~\ref{fig:NS}.

%==========================================================================
%\subsection{Evidence Evaluation}\label{app:nested:evidence}
%==========================================================================

The summation in Eq.~(\ref{eq:NS_sum}) is performed as follows. The iteration counter is first set to~$i=0$ and $N$ `active' (or `live') samples are drawn from the full prior $\pi(\mathbf{\Theta})$, so the initial prior volume is $X_{0} = 1$. The samples are then sorted in order of their likelihood and the smallest (with likelihood $\mathcal{L}_{0}$) is removed from the active set (hence becoming `inactive') and replaced by a point drawn from the prior subject to the constraint that the point has a likelihood $\mathcal{L}>\mathcal{L} _{0}$. The corresponding prior volume contained within the iso-likelihood contour associated with the new live point will be a random variable given by $X_{1} = t_{1} X_{0}$, where $t_{1}$ follows the distribution $\Pr(t) = Nt^{N-1}$ (i.e., the probability distribution for the largest of $N$ samples drawn uniformly from the interval $[0,1]$). At each subsequent iteration $i$, the removal of the lowest likelihood point $\mathcal{L}_{\rm i}$ in the active set, the drawing of a replacement with $\mathcal{L} > \mathcal{L}_{\rm i}$ and the reduction of the corresponding prior volume $X_{\rm i}=t_{\rm i} X_{\rm i-1}$ are repeated, until the entire prior volume has been traversed. The algorithm thus travels through nested shells of likelihood as the prior volume is reduced. The mean and standard deviation of $\log t$, which dominates the geometrical exploration, are: 
\begin{equation}
E[\log t] = -1/N, \quad \sigma[\log t] = 1/N.
\end{equation}
Since each value of $\log t$ is independent, after $i$ iterations the prior volume will shrink down such that $\log X_{\rm i} \approx -(i\pm\sqrt{i})/N$. Thus, one takes $X_{\rm i} = \exp(-i/N)$.

%==========================================================================
%\subsection{Stopping Criterion}\label{nested:stopping}
%==========================================================================

The nested sampling algorithm is terminated when the evidence has been computed to a pre-specified precision. The evidence that could be contributed by the remaining live points is estimated as $\Delta{\mathcal{Z}}_{\rm i} = \mathcal{L}_{\rm max}X_{\rm i}$, where ${\cal L}_{\rm max}$ is the maximum-likelihood value of the remaining live ppoints, and $X_{\rm i}$ is the remaining prior volume. The algorithm terminates when $\Delta{\mathcal{Z}}_{\rm i}$ is less than a user-defined value (we use $0.5$ in log-evidence).

%==========================================================================
%\subsection{Posterior Inferences}\label{nested:posterior}
%==========================================================================

Once the evidence~$\mathcal{Z}$ is found, posterior inferences can be easily  generated using the final live points and the full sequence of discarded points from the nested sampling  process, i.e., the points with the lowest likelihood value at each iteration~$i$ of  the algorithm. Each such point is simply assigned the probability weight
\begin{equation}
p_{\rm i}=\frac{\mathcal{L}_{\rm i}w_{\rm i}}{\mathcal{Z}}.
\label{eq:12}
\end{equation} 
These samples can then be used to calculate inferences of posterior parameters such as  means, standard deviations, covariances and so on, or to construct marginalised posterior distributions.

%==========================================================================
\subsection{The {\sc MultiNest} Algorithm}\label{sec:method:bayesian:multinest}
%==========================================================================

The most challenging task in implementing nested sampling is to draw samples from the prior within the hard constraint $\mathcal{L}> \mathcal{L}_{\rm i}$ at each iteration $i$. The {\sc MultiNest} algorithm \cite{feroz08,multinest} tackles this problem through an ellipsoidal rejection sampling scheme. The  live point set is enclosed within a set of (possibly overlapping)  ellipsoids and a new point is then drawn uniformly from the region enclosed by these ellipsoids. The ellipsoidal decomposition of the live point set is chosen to minimize the sum of volumes of the ellipsoids. The ellipsoidal decomposition is well suited to dealing with posteriors that have curving degeneracies, and allows mode identification in multi-modal posteriors. If there are subsets of the ellipsoid set that do not overlap with the remaining ellipsoids, these are identified as a distinct mode and subsequently evolved independently.

The {\sc MultiNest} algorithm has proven very useful for tackling inference problems in cosmology and particle physics (see e.g. \citealt{2009arXiv0911.0976S, 2009MNRAS.400.1075B, 2009arXiv0910.0760V, 2009PhRvD..79l3521S, 2009arXiv0904.2548A, 2009PhRvD..80c5017A, 2009MNRAS.398.2049F, 2008arXiv0810.0781F, 2009arXiv0903.2487F, 2008JHEP...10..064F, 2008JHEP...12..024T, 2009PhRvD..80i5013L, 2009arXiv0911.1986R,Trotta:2009gr,Roszkowski:2009ye}) typically showing two orders of magnitude improvement in efficiency over conventional techniques. More recently, it has been shown to perform well as a search tool for gravitational wave data analysis \citep{2009CQGra..26u5003F, 2009arXiv0911.0288F}.

%%%%%%%%%%%%%%%%%%%%%%%%%%%%%%%%%%%%%%%%%%%%%%%%%%%%%%%%%
\section{Detailed Critique of van Haasteren (2009)}\label{sec:critique}
%%%%%%%%%%%%%%%%%%%%%%%%%%%%%%%%%%%%%%%%%%%%%%%%%%%%%%%%%

In \cite{vanHaasteren:2009yg}, the author proposes a method based on Vornoi tessellation to calculate the Bayesian evidence (Eq.~\ref{eq:Z}) as follows:
\begin{equation}
\mathcal{Z} = {\textstyle {\displaystyle \sum_{\rm i}}\mathcal{L}_{\rm i}O_{\rm i}},
\label{eq:Z_Haasteren}
\end{equation}
where the index $\rm i$ iterates over the samples (may be obtained through an MCMC algorithm) in the parameter space and $O_{\rm i}$ is the area of the Vornoi cell occupied by the $i^{th}$ sample. Although this approximation converges to the true evidence value, it can not be used in practice because of the high computational cost involved in the calculation of Vornoi tesselation. In order to overcome this problem, the author suggests to concentrate on a small subset of samples $F_{\rm t}$, around the peak of the distribution, occupying volume $V_{\rm t}$ and exploits the following relationship:
\begin{equation}
V_{\rm t} = {\textstyle {\displaystyle \sum_{\mathbf{\Theta}_{\rm i} \in F_{\rm t}}}\frac{\alpha}{\mathcal{L}_{\rm i}}},
\label{eq:alpha}
\end{equation}
where $\alpha$ is a proportionality constant to be determined. In order to calculate $\alpha$, the author then suggests to make a Gaussian approximation around the peak so that the volume $V_{\rm t}$ can be calculated as,
\begin{equation}
V_{\rm t} = \frac{r^{D} \pi^{D/2}}{\Gamma(1 + \frac{D}{2})} \sqrt{|\mathbf{C}|},
\label{eq:Haasteren_ellipse}
\end{equation}
where $D$ is the dimensionality of the problem, $\mathbf{C}$ the covariance matrix and $r$ is the Mahalanobis distance from the peak of the point with the lowest likelihood value inside the region $F_{\rm t}$. Once $\alpha$ has been calculated using Eqs.\eqref{eq:alpha} and \eqref{eq:Haasteren_ellipse}, The evidence can be calculated as:
\begin{equation}
\mathcal{Z} = N \alpha,
\label{eq:Z_Haasteren2}
\end{equation}
where $N$ is the total number samples inside the region $F_{\rm t}$. The author then applies this method to a few toy problems and compares it with {\sc MultiNest}, attempting to show that it outperforms {\sc MultiNest} in terms of accuracy as well as computational efficiency.

We believe that the comparison of {\sc MultiNest} to the method proposed in \cite{vanHaasteren:2009yg} is completely unfair, as it makes a very strong assumption about the availability of suitable posterior samples. Moreover, the criticism of nested sampling algorithm is due to author's misunderstanding about the method. We discuss these in detail now.

%==========================================================================
\subsection{Nested Sampling solves the full inference problem}\label{sec:critique:apples}
%==========================================================================

The method proposed in \cite{vanHaasteren:2009yg} and briefly summarized in the previous section requires the region $F_{\rm t}$ to be sufficiently and adequately populated by MCMC samples but does not propose an MCMC algorithm to satisfy this condition. Therefore, the proposed method does not provide a complete solution for Bayesian inference problem, but it actually makes the strong assumption that such a solution has already been implemented using another technique. For many problems of interest (e.g. problem exhibiting strong, curving degeneracies and/or multi-modality) adequately sampling the parameter space is a difficult problem which often proves to be a stumbling block even for the parameter inference step. The algorithm proposed by \cite{vanHaasteren:2009yg}  assumes that this part of the problem has already been solved, which strongly limits the usefulness of the suggested method if no efficient way of gathering posterior samples is put forward.  

%The difficulty faced by MCMC based algorithms in exploring highly multi--modal problems as well as uni--modal problems with curving degeneracies between the parameters, is well known (see e.g. \citealt{MacKay}) and is an active area of research. Assuming the presence of a perfect MCMC algorithm is equivalent to assuming that one can sample perfectly from the prior within the hard constraint $\mathcal{L}> \mathcal{L}_{\rm i}$ at each iteration $i$ of the nested sampling algorithm. 

The {\sc MultiNest} algorithm, on the other hand, provides a means to carry out both parameter exploration and the evidence evaluation in an efficient and robust way. This has been widely demonstrated by applying it very successfully to various real inference problems in astrophysics, cosmology and particle physics phenomenology (see e.g. \citealt{2009arXiv0911.0976S, 2009MNRAS.400.1075B, 2009arXiv0910.0760V, 2009PhRvD..79l3521S, 2009arXiv0904.2548A, 2009PhRvD..80c5017A, 2009MNRAS.398.2049F, 2008arXiv0810.0781F, 2009arXiv0903.2487F, 2008JHEP...10..064F, 2009CQGra..26u5003F, 2009arXiv0911.0288F, 2008JHEP...12..024T, 2009PhRvD..80i5013L, 2009arXiv0911.1986R}). 

%Throughout his paper, \cite{vanHaasteren:2009yg} assumes that the parameter space has been sampled perfectly through some MCMC algorithm and then uses the resultant samples to calculate the evidence and compares it with the results obtained through {\sc MultiNest}. 
The eggbox problem (discussed in Sec. 5.2 of \citealt{vanHaasteren:2009yg}) is a particular example of what we believe is an unfair comparison. The author assumes that a suitable trick can be found enabling the MCMC algorithm to explore all the modes adequately so that his proposed method can be applied to this highly multi--modal problem, while {\sc MultiNest} not only finds all the modes without making any assumptions but also calculates the evidence accurately.

A fair comparison in our opinion would require the specification of another algorithm capable of providing not only the means to calculate the evidence but also to explore the parameter space without making too many assumptions. In our view, the comparison between the method of  \cite{vanHaasteren:2009yg} and  {\sc MultiNest} is fundamentally unfair, as the latter has a much wider applicability and solves the full inference problem from scratch. 

One of the most attractive features of {\sc MultiNest} is its general applicability for moderately dimensional but highly complex problems. It is completely application--independent in the sense that the specific problem being tackled enters only through the likelihood computation, and does not change how the live point set is updated. It makes no assumptions about the number and nature of modes. Even if a perfect MCMC sampler were available, the method proposed in \cite{vanHaasteren:2009yg} although can be quite useful for uni--modal Gaussian problems, would nonetheless fail for highly non--Gaussian problems. The Gaussian--shell problem discussed in Sec. 6.2 of \cite{multinest} would perhaps be the most obvious example. Comparing {\sc MultiNest} with the proposed method on problems ideally suited to the proposed method is another reason why we think the comparison is completely unfair.

%==========================================================================
\subsection{Misconceptions about Nested Sampling}\label{sec:critique:nested_sampling}
%==========================================================================

In the introduction, the author argues that {\sc MultiNest} (and consequently nested sampling) algorithm by design, samples from the prior distribution and not the posterior distribution, and consequently suffers more with the `curse of dimensionality' than the traditional MCMC based algorithms. This statement ignores the fact that nested sampling does not sample from the prior distribution blindly, it samples from the prior within the hard constraint $\mathcal{L}> \mathcal{L}_{\rm i}$ at each iteration $i$. As discussed in Sec.~\ref{sec:nested}, this hard--edge sampling scheme results in exponential decrease in the prior volume $X_{\rm i}$ occupied by the live points at the $i^{th}$ iteration with the expected value of $X_{\rm i}$ given as,
\begin{equation}
<X_{\rm i}> = \exp(-i/N),
\end{equation}
$N$ being the total number of live points. This allows the algorithm to reach highly localized regions of the parameter space with large likelihood values in a reasonable number of iterations. We can not see why such a sampling scheme would suffer from the `curse of dimensionality' more than the traditional MCMC schemes. 

In our opinion, the information content $H$, given as follows,
\begin{equation}
H = \int \,\log \left(\frac{dP}{dX}\right)\,dX,
\label{eq:info}
\end{equation}
where $P$ denotes the posterior, is more important than the dimensionality of the problem. Most of the contribution to the evidence value usually comes from the iterations around the maximum likelihood point, which occurs in the region with prior volume $X \approx e^{-H}$ and therefore because of the exponential shrinkage of the prior volume, one could argue that nested sampling has an inherent advantage over MCMC schemes. It should be noted that nested sampling framework itself leaves it to the user to design a scheme to sample from the iso--likelihood contour and it is certainly possible to come up with highly inefficient schemes, to sample a new point with $\mathcal{L}> \mathcal{L}_{\rm i}$, suffering severely with the `curse of dimensionality'.

In Sec. 4.2, the author claims that nested sampling generates samples from the whole of the parameter space, rather than from the posterior distribution, and therefore it can never reach the efficiency achieved by the traditional MCMC based methods. This statement ignores two key points. First, as discussed earlier in this section, nested sampling does not sample from the whole of the prior distribution, but instead it samples from the hard--edge region inside the prior whose volume is reduced exponentially with each iteration. Secondly, the samples generated by a nested sampling algorithm are not all equally weighted as the ones generated through an MCMC method are. As discussed in Sec.~\ref{sec:nested}, one needs to assign a probability weight given by Eq.~\eqref{eq:12} to the point with lowest likelihood value at each iteration of the nested sampling algorithm. In Fig. 3 of \cite{vanHaasteren:2009yg}, the author shows a scatter plot of 40,000 samples obtained through a nested sampling algorithm. The author does not mention whether he has plotted simply the points with the lowest likelihood values at each iteration or the probability weights of these points have also been taken into account and therefore we are unable to comment on the accuracy of the figure.

%==========================================================================
\subsection{Curse of Dimensionality}\label{sec:critique:dimensionality}
%==========================================================================

The author repeatedly attacks nested sampling algorithm in general and {\sc MultiNest} in particular, saying that it suffers severely from the `curse of dimensionality'. We have already discussed in the previous section that as far as the general nested sampling framework is concerned, this is not true. 

{\sc MultiNest} implements the nested sampling algorithm through an ellipsoidal rejection sampling scheme to sample uniformly from the iso--likelihood contour as discussed in Sec.~\ref{sec:method:bayesian:multinest}. It is well known that all rejection sampling schemes are highly inefficient for high dimensional problems (see e.g. \citealt{MacKay}). {\sc MultiNest} was designed to work with problems with moderately high number of dimensions and it has proven highly successful to deal with highly multi--modal and complex problems in cosmology and particle physics phenomenology. It should be noted that nested sampling not only provides a way to evaluate the Bayesian evidence accurately but with a clever algorithm to sample from the hard constraint, it also provides a solution to deal with highly degenerate and multi--modal problems. A very good example is the use of {\sc MultiNest} in gravitational wave astronomy (see e.g. \citealt{2009CQGra..26u5003F, 2009arXiv0911.0288F}) where the problems are inherently multi--modal and so far {\sc MultiNest} has mainly been used as a parameter exploration tool with great deal of success. 

For parameter exploration, {\sc MultiNest} works with reasonable efficiency up to $\sim$ 100D beyond which the efficiency drops appreciably as expected. Accurate evidence evaluation requires adequate sampling from the whole of parameter space and even slight inaccuracies in estimating the iso--likelihood region through the ellipsoidal decomposition can result in large inaccuracies and therefore {\sc MultiNest} can calculate the evidence value accurately with reasonable efficiency for problems up to $\sim$ 50D. In order to demonstrate this, we ran {\sc MultiNest} on an $n$--dimensional ellipsoidal Gaussian with likelihood \footnote{We would have liked to test {\sc MultiNest} on the same toy problem described in Sec. 5.1 of \cite{vanHaasteren:2009yg} but we were unable to do so as the author did not describe the prior distribution he used and hence we chose a similar but not exactly the same problem.},
\begin{equation}
\mathcal{L} = \displaystyle\prod_{\rm i = 1}^{\rm n} \frac{1}{\sqrt{2 \pi \sigma_{\rm i}}} \exp \left( -\frac{\theta_{\rm i} - 0.5}{2 \sigma_{\rm i}} \right),
\label{eq:nD_gaussian}
\end{equation}
with
\begin{equation}
\sigma_{\rm i} = 0.001 {\rm i}.
\label{eq:nD_gaussian_sigma}
\end{equation}
We set uniform prior $\mathcal{U}(0,1)$ for all the parameters so that the analytical $\log$-evidence value is 0.0 regardless of the dimensionality of the problem. We used 1,000 live points with target efficiency $e$ set to 1.0 for the 2D problem and reducing it to 0.01 for the 32D problem. A value of $e = 0.3$ was suggested in \cite{multinest} for the standard CMB data analysis, but it was accompanied by the caveat that the user should check that the evidence value is consistent when $e$ is lowered. We list the number of likelihood evaluations, recovered $\log(\mathcal{Z})$ and information content $H$ in Table ~\ref{tab:gaussian}. These results clearly show that {\sc MultiNest} is able to correctly evaluate the evidence values even for the $32D$ problem with the posterior occupying only $e^{-96.10}$ of the prior volume, although the efficiency does drop appreciably with the increase in dimensionality of the problem. This drop in efficiency is mainly due to the exponential increase in the information content.
\begin{table}
\begin{center}
\begin{tabular}{rrrr}
\hline
$n$ & $N_{\rm like}$ & $\log(\mathcal{Z})$  & $H$ \\
\hline
$2$   & $14,184$ & $0.04 \pm 0.10$ & $10.00$ \\
$4$   & $26,033$ & $-0.07 \pm 0.14$ & $19.60$ \\
$8$   & $107,876$ & $-0.24 \pm 0.18$ & $32.40$ \\
$16$  & $651,345$ & $0.11 \pm 0.24$ & $57.60$ \\
$32$  & $14,134,227$ & $0.40 \pm 0.31$ & $96.10$ \\
\hline
\end{tabular}
\caption{The log--evidence ($\log(\mathcal{Z})$) and information content ($H$) values obtained by the {\sc MultiNest} algorithm when applied to the problem described in Eq.~\eqref{eq:nD_gaussian}. The analytical $\log(\mathcal{Z})$ is 0.0 regardless of dimensionality $n$.}
\label{tab:gaussian}
\end{center}
\end{table}

We should also mention that for a Gaussian problem, we have been able to calculate the evidence value accurately up to $\sim$ 1000D using nested sampling with a Hamiltonian sampling scheme to sample from the hard--constraint. This method is still under development and we would present the results in a forthcoming publication.

%==========================================================================
%\subsection{General Applicability}\label{sec:critique:general_applicability}
%==========================================================================

%%%%%%%%%%%%%%%%%%%%%%%%%%%%%%%%%%%%%%%%%%%%%%%%%%%%%%%%%
\section{Conclusions}\label{sec:conclusions}
%%%%%%%%%%%%%%%%%%%%%%%%%%%%%%%%%%%%%%%%%%%%%%%%%%%%%%%%%

{\sc MultiNest} has proven to be a very useful and powerful tool to carry out Bayesian inference for a wide variety of problems in cosmology and particle physics phenomenology as well as for general moderately dimensional inference problems. While the method proposed in \citealt{vanHaasteren:2009yg} to calculate the Bayesian evidence is interesting for sufficiently simple problems, it does not have general applicability and relies on the availability of samples distributed according to the posterior distribution. The comparison of this method with {\sc MultiNest} is completely unfair as {\sc MultiNest} provides the means to perform full Bayesian analysis without making any assumptions about the nature of the problem nor does it rely on the availability of posterior samples, in fact it can be used to provide the posterior samples.

%\begin{thebibliography}{99}
\bibliographystyle{mn2e}
\bibliography{references}

\begin{thebibliography}{}

\bibitem[\protect\citeauthoryear{AbdusSalam, Allanach, Dolan, Feroz \&
  Hobson}{AbdusSalam et~al.}{2009a}]{2009PhRvD..80c5017A}
AbdusSalam S.~S.,  Allanach B.~C.,  Dolan M.~J.,  Feroz F.,    Hobson M.~P.,
  2009a, \prd, 80, 035017 [arXiv:0906.0957]

\bibitem[\protect\citeauthoryear{AbdusSalam, Allanach, Quevedo, Feroz \&
  Hobson}{AbdusSalam et~al.}{2009b}]{2009arXiv0904.2548A}
AbdusSalam S.~S.,  Allanach B.~C.,  Quevedo F.,  Feroz F.,    Hobson M.,  2009b,
  ArXiv e-prints [arXiv:0904.2548]

\bibitem[\protect\citeauthoryear{Bridges, Feroz, Hobson \& Lasenby}{Bridges
  et~al.}{2009}]{2009MNRAS.400.1075B}
Bridges M.,  Feroz F.,  Hobson M.~P.,    Lasenby A.~N.,  2009, \mnras, 400,
  1075 [arXiv:0812.3541]

\bibitem[\protect\citeauthoryear{{Feroz}, {Allanach}, {Hobson}, {Abdus Salam},
  {Trotta} \& {Weber}}{{Feroz} et~al.}{2008a}]{2008JHEP...10..064F}
{Feroz} F.,  {Allanach} B.~C.,  {Hobson} M.,  {Abdus Salam} S.~S.,  {Trotta}
  R.,    {Weber} A.~M.,  2008a, Journal of High Energy Physics, 10, 64 [arXiv:0807.4512]

\bibitem[\protect\citeauthoryear{Feroz, Gair, Graff, Hobson \& Lasenby}{Feroz
  et~al.}{2009a}]{2009arXiv0911.0288F}
Feroz F.,  Gair J.~R.,  Graff P.,  Hobson M.~P.,    Lasenby A.,  2009a, ArXiv
  e-prints [arXiv:0911.0288]

\bibitem[\protect\citeauthoryear{Feroz, Gair, Hobson \& Porter}{Feroz
  et~al.}{2009b}]{2009CQGra..26u5003F}
Feroz F.,  Gair J.~R.,  Hobson M.~P.,    Porter E.~K.,  2009b, Classical and
  Quantum Gravity, 26, 215003 [arXiv:0904.1544]

\bibitem[\protect\citeauthoryear{Feroz \& Hobson}{Feroz \&
  Hobson}{2008}]{feroz08}
Feroz F.,  Hobson M.~P.,  2008, \mnras, 384, 449 [arXiv:0704.3704]

\bibitem[\protect\citeauthoryear{Feroz, Hobson \& Bridges}{Feroz
  et~al.}{2009c}]{multinest}
Feroz F.,  Hobson M.~P.,    Bridges M.,  2009c, \mnras, 398, 1601 [arXiv:0809.3437]

\bibitem[\protect\citeauthoryear{Feroz, Hobson, Roszkowski, Ruiz~de Austri \&
  Trotta}{Feroz et~al.}{2009d}]{2009arXiv0903.2487F}
Feroz F.,  Hobson M.~P.,  Roszkowski L.,  Ruiz~de Austri R.,    Trotta R.,
  2009d, ArXiv e-prints [arXiv:0903.2487]

\bibitem[\protect\citeauthoryear{Feroz, Hobson, Zwart, Saunders \&
  Grainge}{Feroz et~al.}{2009e}]{2009MNRAS.398.2049F}
Feroz F.,  Hobson M.~P.,  Zwart J.~T.~L.,  Saunders R.~D.~E.,    Grainge
  K.~J.~B.,  2009e, \mnras, 398, 2049 [arXiv:0811.1199]

\bibitem[\protect\citeauthoryear{Feroz, Marshall \& Hobson}{Feroz
  et~al.}{2008}]{2008arXiv0810.0781F}
Feroz F.,  Marshall P.~J., Hobson M.~P.,  2008, ArXiv e-prints [arXiv:0810.0781]

\bibitem[\protect\citeauthoryear{{Hobson}, {Bridle} \& {Lahav}}{{Hobson}
  et~al.}{2002}]{Hobson02}
{Hobson} M.~P.,  {Bridle} S.~L., {Lahav} O.,  2002, \mnras, 335, 377 [arXiv:astro-ph/0203259]

\bibitem[\protect\citeauthoryear{Kilbinger et~al.,}{Kilbinger
  et~al.}{2009}]{Kilbinger:2009by}
Kilbinger M.,  et~al., 2009, ArXiv e-prints [arXiv:0912.1614]

\bibitem[\protect\citeauthoryear{Liddle}{Liddle}{2004}]{Liddle:2004nh}
Liddle A.~R.,  2004, Mon. Not. Roy. Astron. Soc., 351, L49 [arXiv:astro-ph/0401198]

\bibitem[\protect\citeauthoryear{{Liddle}}{{Liddle}}{2007}]{Liddle07}
{Liddle} A.~R.,  2007, Mon. Not. Roy. Astron. Soc., 377, L74 [arXiv:astro-ph/0701113]

\bibitem[\protect\citeauthoryear{{L{\'o}pez-Fogliani}, {Roszkowski}, {de
  Austri} \& {Varley}}{{L{\'o}pez-Fogliani} et~al.}{2009}]{2009PhRvD..80i5013L}
{L{\'o}pez-Fogliani} D.~E.,  {Roszkowski} L.,  {de Austri} R.~R.,    {Varley}
  T.~A.,  2009, \prd, 80, 095013 [arXiv:0906.4911]

\bibitem[\protect\citeauthoryear{{Mackay}}{{Mackay}}{2003}]{MacKay}
{Mackay} D.~J.~C.,  2003, {Information Theory, Inference and Learning
  Algorithms}.
Information Theory, Inference and Learning Algorithms, by David J.~C.~MacKay,
  pp.~640.~ISBN 0521642981.~Cambridge, UK: Cambridge University Press, October
  2003.

\bibitem[\protect\citeauthoryear{{\'O}~Ruanaidh \& Fitzgerald}{{\'O}~Ruanaidh
  \& Fitzgerald}{1996}]{Ruanaidh}
{\'O}~Ruanaidh J.,  Fitzgerald W.,  1996, Numerical Bayesian Methods Applied to
  Signal Processing.
Springer Verlag:New York

\bibitem[\protect\citeauthoryear{{Raklev} \& {White}}{{Raklev} \&
  {White}}{2009}]{2009arXiv0911.1986R}
{Raklev} A.~R.,  {White} M.~J.,  2009, ArXiv e-prints [arXiv:0911.1986]

\bibitem[\protect\citeauthoryear{Roszkowski, Ruiz~de Austri \&
  Trotta}{Roszkowski et~al.}{2009}]{Roszkowski:2009ye}
Roszkowski L.,  Ruiz~de Austri R.,    Trotta R.,  2009, ArXiv e-prints [arXiv:0907.0594]

\bibitem[\protect\citeauthoryear{{Sekiguchi}, {Ichikawa}, {Takahashi} \&
  {Greenhill}}{{Sekiguchi} et~al.}{2009}]{2009arXiv0911.0976S}
{Sekiguchi} T.,  {Ichikawa} K.,  {Takahashi} T.,    {Greenhill} L.,  2009,
  ArXiv e-prints [arXiv:0911.0976]

\bibitem[\protect\citeauthoryear{{Skilling}}{{Skilling}}{2004}]{Skilling04}
{Skilling} J.,  2004, in {Fischer} R.,  {Preuss} R.,   {Toussaint} U.~V.,  eds,
  American Institute of Physics Conference Series {Nested Sampling}.
pp 395--405

\bibitem[\protect\citeauthoryear{{Sollom}, {Challinor} \& {Hobson}}{{Sollom}
  et~al.}{2009}]{2009PhRvD..79l3521S}
{Sollom} I.,  {Challinor} A.,    {Hobson} M.~P.,  2009, \prd, 79, 123521 [arXiv:0903.5257]

\bibitem[\protect\citeauthoryear{Trotta}{Trotta}{2007a}]{Trotta:2005ar}
Trotta R.,  2007a, Mon. Not. Roy. Astron. Soc., 378, 72 [arXiv:astro-ph/0504022]

\bibitem[\protect\citeauthoryear{Trotta}{Trotta}{2007b}]{Trotta:2006ww}
Trotta R.,  2007b, Mon. Not. Roy. Astron. Soc. Lett., 375, L26 [arXiv:astro-ph/0608116]

\bibitem[\protect\citeauthoryear{Trotta}{Trotta}{2008}]{Trotta:2008qt}
Trotta R.,  2008, Contemp. Phys., 49, 71 [arXiv:0803.4089]

\bibitem[\protect\citeauthoryear{Trotta, de Austri \& Heros}{Trotta
  et~al.}{2009}]{Trotta:2009gr}
Trotta R.,  de Austri R.~R.,    Heros C. P. d.~l.,  2009, JCAP, 0908, 034 [arXiv:0906.0366]

\bibitem[\protect\citeauthoryear{{Trotta}, {Feroz}, {Hobson}, {Roszkowski} \&
  {Ruiz de Austri}}{{Trotta} et~al.}{2008}]{2008JHEP...12..024T}
{Trotta} R.,  {Feroz} F.,  {Hobson} M.,  {Roszkowski} L.,    {Ruiz de Austri}
  R.,  2008, Journal of High Energy Physics, 12, 24 [arXiv:0809.3792]

\bibitem[\protect\citeauthoryear{van Haasteren}{van
  Haasteren}{2009}]{vanHaasteren:2009yg}
van Haasteren R.,  2009, ArXiv e-prints [arXiv:0911.2150]

\bibitem[\protect\citeauthoryear{Vardanyan, Trotta \& Silk}{Vardanyan
  et~al.}{2009}]{Vardanyan:2009ft}
Vardanyan M.,  Trotta R.,  Silk J.,  2009, Mon. Not. Roy. Astron. Soc., 397, 431 [arXiv:0901.3354]

\bibitem[\protect\citeauthoryear{{Vegetti}, {Koopmans}, {Bolton}, {Treu} \&
  {Gavazzi}}{{Vegetti} et~al.}{2009}]{2009arXiv0910.0760V}
{Vegetti} S.,  {Koopmans} L.~V.~E.,  {Bolton} A.,  {Treu} T.,    {Gavazzi} R.,
  2009, ArXiv e-prints [arXiv:0910.0760]

\end{thebibliography}
%\end{thebibliography}

\appendix

\label{lastpage}

\end{document}